\documentclass[preprint,preprintnumbers]{revtex4}
\usepackage{amssymb}
\usepackage{graphicx}
\usepackage{bm}
\begin{document}

\preprint{USTC-ICTS-11-02}
\title{The origin of entropy production in spacetime thermodynamics}
\author{Wei Gu}
\email{guwei@mail.ustc.edu.cn}
\affiliation{Interdisciplinary Center for Theoretical Study, University of Science and
Technology of China, Hefei 230026, China}
\affiliation{Department of Modern Physics, University of Science and Technology of China,
Hefei 230026, China}
\author{Rong-Xin Miao}
\email{mrx11@mail.ustc.edu.cn}
\affiliation{Interdisciplinary Center for Theoretical Study, University of Science and
Technology of China, Hefei 230026, China}
\affiliation{Department of Modern Physics, University of Science and Technology of China,
Hefei 230026, China}
\author{Guang Wu}
\email{rocky29@mail.ustc.edu.cn}
\affiliation{Department of Modern Physics, University of Science and Technology of China,
Hefei 230026, China}

\begin{abstract}
We find that the ambiguity term of approximate Killing vector field is
responsible for the entropy production term. Without the ambiguity term,
pure Einstein theory and $f(R)$ satisfy the relation of thermodynamic
equilibrium. Considering such an ambiguity term of approximate Killing
vector field, we can get the entropy production term and the entropy in $f(R)$
with a form defined by Jacobson. In pure Einstein theory, the shear term is
the only geometric contribution of entropy production term, while in $f(R)$ it
can also contribute. We believe our approach and conclusion can be
generalized to other gravity theory.
\end{abstract}

\maketitle


\section{\protect\bigskip Introduction}

\ \ The discovery of black hole entropy and Hawking radiation in 1970s \cite%
{Bekenstein:1973ur} implies a profound connection between gravitation and
thermodynamics. Almost twenty years ago, Jacobson \cite{Jacobson:1995ab}
proposed to interpret this connection by reversing the logic and deriving
the Einstein's equations from a thermodynamic equation of state with two
assumptions: proportionality of entropy and area for all local Rindler
horizons, and the Clausius relation. The first assumption is nothing but the
expression of the holographic principle \cite{Hooft}\cite{Susskind:1994vu} .
In 1998, Maldacena gave a gorgeous conjecture, namely the AdS/CFT
correspondence \cite{Maldacena:1997re}, which is a direct manifestation of
the holographic principle. It was suggested that gravitation is induced by a
quantum field theory in lower dimensions which to a large extent support the
idea of gravitation on the macroscopic scale as a manifestation of the
thermodynamics of the vacuum state. Furthermore, more confidence about gravity
being emergent rather than fundamental was recently inspired in \cite%
{Verlinde:2010hp}\cite{Gu:2010wv}\cite{Padmanabhan:2009vy}\cite{Miao:2011er}\cite{Padmanabhan:2010rp}.

In \cite{Eling:2006aw}, Jacobson treated $f(R)$ theory as non-equilibrium
thermodynamics of spacetime, therefore, an entropy production term is
required to keep the entropy balance relation. The term is an outcome of the
non-vanishing expansion at $p$ which is related to a local boost dependence
quantity. In 2008, Elizalde and Silva \cite{Elizalde and Silva} proposed an
alternative approach, with local thermodynamic equilibrium maintained, using
the idea of \textquotedblleft local-boost-invariance\textquotedblright\
introduced in \cite{Wald:1993nt}. In other words, local boost dependence
quantity is erased under the boost-invariant truncation, so expansion at $p$
vanishes, without any entropy production term when we repeat Jacobson's
derivation. Brustein and Hadad \cite{Brustein:2009hy} showed that the
equations of motion of generalized theories of gravity are equivalent to the
thermodynamics relation $\delta Q=T\delta S$. Their proof relies on
extending previous arguments by using a more general definition,  namely
Noether charge entropy.

In this paper, we focus on the origin and feature of the entropy production.
We show that it is a consequence of the $o(x^{3})$ ambiguity in an
approximate boost Killing vector field. If we suspend this ambiguity first,
the entropy production will vanish, and vice versa. In order to research the
thermodynamics of different gravity models, we assume the validity of those
gravity field equations, and then use them to derive the form of entropy. In
$f(R)$ gravity, if the \textquotedblleft boost variant\textquotedblright (a Lie derivative of $f(R)$ along
Killing vector at a point $p$ in the spacetime) is nonzero, the entropy production is a
function of the \textquotedblleft boost variant\textquotedblright and non-vanishing expansion at $p$. In pure
general relativity (GR), we can not adopt such approach. The shear term from
Raychauduri equation can include the entropy production. $f(R)$ theory can
also accept the shear term, instead of non-vanishing expansion term, as the
entropy production term. At last, we may find that the shear term being the
entropy production term is more general. Conditions of pure GR and $f(R)$
theory analyzed, we believe our method and results can also be generalized
to other gravity theories.

\section{Approximate Killing vector field}

In this section, we review Jacobson's work briefly and set the background of
our work. First, we can describe spacetime in a vicinity of a free-falling
local observer $p$ as flat through the equivalence principle. Then, we
choose a local 2-surface element $B$ including $p$ and perpendicular to the
worldline of $p$. The boundary of the past of $B$ is defined as "local
Rindler horizon" [horizon entropy], whose generators are a congruence of
null geodesics with vanishing expansion and shear. Therefore, the local
Rindler horizon reaches equilibrium at $p$. Considering the local Rindler
horizon is actually a causal horizon, we can introduce an entropy $S$,
measuring the degrees of freedom beyond it. According to the holographic
principle, $S$ is proportionate to the area elements of the horizon.

\ The definition of the heat flux and temperature is related to an
approximate boost Killing vector field $\chi ^{a}$. $\chi ^{a}$ generating
boosts orthogonal to $B$ the causal horizon. It vanishes at $p$, its flow
invariant at the tangent plane $B_{p}$. Its covariant derivative $\chi
_{a;b} $ is a timelike antisymmetric tensor orthogonal to $B_{p}$. We
normalize $\chi ^{a}$ by $\chi _{a;b}\chi ^{a;b}=-2$. In a common curved
spacetime, no Killing vectors exist. we can only solve the Killing equation $%
\nabla ^{a}\chi ^{b}+\nabla ^{b}\chi ^{a}=0$ with this "initial data" out to
some order in the neighborhood of $p$. The equation
\begin{equation}
\nabla _{a}\nabla _{b}\chi ^{a}=R_{ab}\chi ^{a},  \label{Eq1}
\end{equation}%
is equal to the Killing equation. In Riemann normal coordinates $\{e_{\mu
}^{a}\}$ based at $p$, the zeroth and the first order parts of $\chi ^{a}$
are resolved by initial conditions, while the second order part vanishes
according to the Killing's equation (compare both sides of Eq.\ref{Eq1} to
get this conclusion). Generally, the equation cannot be satisfied at third
order, so we still have a $o(x^{3})$ ambiguity in Killing vector $\chi ^{a}$%
, which would influence the integrability of equation \ref{Eq1}(refer to
Weinberg's text book \cite{Weinberg} for detail). We choose the direction of
$\chi ^{a}$ to be future pointing on the causal horizon. In flat spacetime, $%
\chi ^{a}$ can be defined as $\chi ^{a}=-\lambda k^{a}$, where $k^{a}$ is
the horizon tangent vector and $\lambda $ is a negative affine parameter
that is increasing along the horizon and vanishes at $p$. Here, we set
light-cone coordinates as $k^{a}=(e_{0}^{a}+e_{3}^{a})/\sqrt{2}%
,l^{a}=(e_{0}^{a}-e_{3}^{a})/\sqrt{2},x^{a}=e_{1}^{a},y^{a}=e_{2}^{a},$
which satisfy
\begin{equation}
g_{ab}k^{a}l^{b}=-1,\ \ \ g_{ab}k^{a}k^{b}=g_{ab}l^{a}l^{b}=0.   \label{2}
\end{equation}%
\qquad With the third order ambiguity in curved spacetime, $\chi ^{a}$
should be
\begin{equation}
\chi ^{a}=-\lambda k^{a}+o(x^{3}) .  \label{3}
\end{equation}%
and therefore Eq.\ref{Eq1} becomes
\begin{equation}
\nabla _{a}\nabla _{b}\chi ^{a}=R_{ab}\chi ^{a}+o(x) .  \label{4}
\end{equation}

After defining the local Rindler horizon and approximate Killing vectors, it
is natural to define the heat. The heat is the mean flux of the boost energy
current across the horizon measured by a uniformly accelerated observer
hovering inside the horizon:
\begin{equation}
\delta Q=\int_{H}T_{ab}\chi ^{a}\epsilon ^{b}
\end{equation}%
$T_{ab}$ is the expectation value of the matter stress tensor, and $\epsilon
^{b}$ is the area of each cross section element of $H$. The integration is
over a short pencil of horizon generators of $H$, and $\lambda $ contained
in $\chi ^{a}$ represents the evolution of those generators. Since it is an
equilibrium state at $p$, the metric and, further, its conjugate $T_{ab}$
are also approximately stable. It means the expectation value of the matter
stress tensor do not change over a sufficiently small $\lambda $ or,
equivalently, quantum transitions terminate. Without loss of generality, we
choose $T_{ab}|_{p}=T_{ab}|_{\lambda =0}$.

According to the Unruh effect and Rindler coordinates, we know acceleration
tends to diverge as observer approaches the horizon. Since both of the
temperature and heat flux are proportional to the acceleration, they diverge
the same rate. Thus, we can choose the acceleration to be unit and the
temperature measured by the observer to be $T=\hbar /2\pi $.

\section{\protect\bigskip The calculation of f(R)}

In Jacobson's paper, he reversed the logic, using two hypotheses, universal
entropy density of the horizon and the Clausius relation in vicinity of $B$,
to derive the Einstein equation. Some gravity theories, like $f(R)$ gravity,
may need an entropy production rate to keep the entropy balance relation.
However, we wonder that the entropy may contain un-physical quantities which
will cause the entropy production. In \cite{Elizalde and Silva}\cite%
{Brustein:2009hy}, Wald entropy was introduced immediately, so the
un-physical quantities were truncated.

In this section, we postulate the $f(R)$ gravity field equation first, and
then ascertain the form of entropy affecting the dynamics. Considering the
Killing equation and approximate Killing vector field, we will find the
entropy production rate arising from the ambiguity term.

The equation of motion of $f(R)$ is
\begin{equation}
F(R)R_{ab}(g)-\frac{1}{2}f(R)g_{ab}-\nabla _{a}\nabla _{b}F(R)+g_{ab}\square
F(R)=8\pi GT_{ab}.
\end{equation}
Put it into the heat, it follows that
\begin{equation}
\delta Q=\frac{1}{8\pi G}\int \left( FR_{ab}-\nabla _{a}\nabla _{b}F\right)
\chi ^{a}k^{b}d\lambda dA .  \label{Eq7}
\end{equation}%

For simplicity, we leave the ambiguity term $o(x^{3})$ of Eq.\ref{Eq1}
alone and pick it up afterward. Terms with $g_{ab}$ vanish since $g_{ab}$
contracted with $\chi ^{a}k^{b}$ leads to $-\lambda k_{a}k^{a}=0$ in
light-cone coordinates. We then evaluate the equation at the leading terms
in $\lambda $,
\begin{equation}
\frac{\delta Q}{T}=\frac{1}{4\hbar G}\int \left( F|_{p}R_{ab}-\nabla
_{a}\nabla _{b}F|_{p}\right) \chi ^{a}k^{b}d\lambda dA=\frac{1}{4\hbar G}%
\int (F|_{p}\nabla _{a}\nabla _{b}\chi ^{a}-\nabla _{a}\nabla _{b}F|_{p}\chi
^{a})k^{b}d\lambda dA.
\end{equation}%
The Killing equation shows $\nabla _{b}\chi _{a}$ is antisymmetric, so we
can use Stoke's theorem and obtain
\begin{equation}
\frac{\delta Q}{T}=\frac{1}{4\hbar G}\{\int l_{a}F|_{p}k^{b}(\nabla _{b}\chi
^{a})dA|_{0}^{d\lambda }+\int \lambda \ddot{F}|_{p}d\lambda dA\}=\frac{1}{%
4\hbar G}\int (\theta F|_{p}+\lambda \ddot{F}|_{p})d\lambda dA.  \label{Eq9}
\end{equation}%
Here comes a contradiction that the integrand contain $\lambda $ at
beginning, while the first term at the last step do not. In other words, $%
\theta F|_{p}$ has a term of less order $\theta F|_{p}=\theta|
_{p}F|_{p}+\lambda \dot{\theta}|_{p}F|_{p}.$ It seems that $\theta|_{p}$
should be zero by comparing the two sides's order of Eq.\ref{Eq9},
therefore $\frac{\delta Q}{T}=\frac{1}{4\hbar G}\int (\lambda \dot{\theta}%
|_{p}F|_{p}+\lambda \ddot{F}|_{p})$

Let's check the value of $\theta| _{p}$ in another way. In Jacobson's paper,
the entropy change is $\delta S=\alpha \int (\theta F+\dot{F})d\lambda dA$,
where $\alpha =\frac{1}{4\hbar G}$. It can be extracted as
\begin{equation}
\delta S=\alpha \int [(\lambda \dot{\theta}|_{p}F|_{p}+\lambda \ddot{F}%
|_{p})+(\lambda \theta|_{p}\dot{F}|_{p}+\theta|_{p}F|_{p}+\dot{F}%
|_{p})]d\lambda dA.
\end{equation}
and the entropy balance relation is $dS=\delta Q/T+d_{i}S$, $\dot{F}\mid_{p}$ is the \textquotedblleft boost variant\textquotedblright which mentioned in the introduction. The Raychaudhuri equation is
\begin{equation}
\frac{d\theta }{d\lambda }=-\frac{1}{2}\theta ^{2}-\sigma ^{ab}\sigma
_{ab}-R_{ab}k^{a}k^{b} .  \label{Eq11}
\end{equation}%
the shear term is required to be zero at $p$. Comparing these equations, we
will obtain the entropy production term
\[
d_{i}S=\alpha \int (\lambda \theta|_{p}\dot{F}|_{p}-\frac{\lambda}{2}%
\theta^{2}|_{p}+\theta|_{p}F|_{p}+\dot{F}|_{p})d\lambda dA.
\]
However, the entropy production term is supposed to vanish at $p$, the
equilibrium point, so the rate should be of order $\lambda $. It requires $%
\theta| _{p}F|_{p}+\dot{F}|_{p}=0$ and $\theta|_{p}=0$

\begin{equation}
\frac{\delta Q}{T}=\alpha \int (\lambda \dot{\theta}|_{p}F|_{p}+\lambda
\ddot{F}|_{p})d\lambda dA.
\end{equation}%
This equation is directly related to the equation of motion of $f(R).$

Now, if we use $\chi ^{a}=-\lambda k^{a}+o(x^{3})$ in Eq.%
\ref{Eq7}, then the integrand will have $o(x)$ $\thicksim \nabla \nabla
o(x^{3})$ which has the same order with other terms. $o(x)$ $\thicksim
\nabla \nabla o(x^{3})$ consist of two parts: one is from Eq.\ref{4}, the
other is from using the Stokes theorem. The ambiguity term $o(x)$ act as $%
\lambda \theta|_{p}\dot{F}|_{p},$ which also vanishes at $p$ and is
irrelevant to the equation of motion. Substituting $\dot{F}%
|_{p}=-\theta|_{p}F|_{p}$ to this term, we will get $-\alpha \lambda
\theta|_{p}^{2}F|_{p}$ parallel with the entropy production density derived
by Jacobson.

One thing need attention is we only use $\theta|_{p}F|_{p}+\dot{F}|
_{p}=0$ in this part, not each of these two terms to be zero. It means that
they are not related to the the equation of motion. $\dot{F}%
|_{p}=F^{^{\prime }}(R)k^{a}R_{,a}$ is a ``boost variant'', which is depandant
on the choice of coordinates and so is $\theta|_{p}$ which is restricted by $%
\dot{F}|_{p}.$ Hence, we find the entropy production is related to boost.

\section{Compare with GR}

In this section, we discuss the situation of pure GR
under a similar procedure. Replace the equation of motion of $f(R)$ with
Einstein equation $R_{ab}-\frac{1}{2}Rg_{ab}=T_{ab},$ we immediately get
\begin{equation}
\frac{\delta Q}{T}=\frac{1}{4\hbar G}\int R_{ab}\chi ^{a}k^{b}d\lambda dA=%
\frac{1}{4\hbar G}\int (\theta +o(x))d\lambda dA
\end{equation}%
here $\theta| _{p}=0$ is demanded for the reason that both sides of the
equation should have the same order. Without $o(x),$ we can get the state
equation of equilibrium positively. However, if we consider ambiguity term,
we can not formulate the entropy production term since expansion at $p$
vanishies. However, in Raychaudhuri equation (Eq.\ref{Eq11}) there is a
shear term $\sigma ^{2}=\sigma ^{ab}\sigma _{ab}$. If the shear term is not
required to be zero, it can be used to describe the entropy production term
such as $\alpha \sigma ^{ab}\sigma _{ab}$. This is coincident with
Jacobson's remarks. Actually, entropy production term comprising $\theta
|_{p}$ is not indispensable, although $\theta $\ can be limited by $\dot{f}%
|_{p}$. If, in $f(R)$ gravity, we employ the condition of nonzero shear like
in pure GR, we will also keep $\theta| _{p}=0$. It may, from another aspect,
give the reason why the thermodynamic relation $\delta Q=T\delta S$ hold in
generalized theories of gravity.

\section{Conclusion}

\bigskip These two situations analyzed above indicate that the entropy
production term comes from the ambiguity term of the approximate Killing
vector. We do need it for the entropy balance relation rather than the field
equation. From this point of view, we find the entropy production term is
not related to the equation of motion. This ambiguity term can also be constructed of
nonvanishing $\sigma ^{2}$ or other quantities. It can be seen that our
method is universal, therefore we make the conclusion that entropy
production term is caused by the ambiguity term of the approximate Killing
vector field. Furthermore, we believe that the entropy production term can
also be described by nonvanishing $\sigma ^{2}$, since Raychaudhuri equation
is used generally on null geodesics distortion caused by gravity.

\section{Acknowledgments}

We thank  Miao Li and Xiao-Dong Li  for valuable discussions and suggestions.

\end{document}